\documentclass{iau} 

\newcommand{\atlas}{ATLAS$^{3\mathrm{D}}$}

\usepackage{graphicx}
\usepackage{xcolor}
\usepackage[hyphens]{url}
\usepackage{hyperref}
\hypersetup{
     colorlinks  = true,
     linkcolor   = blue, 
     anchorcolor = BrickRed,
     citecolor   = blue,
     filecolor   = blue,
     menucolor   = blue,
     runcolor    = cyan,
     urlcolor    = blue
}
\usepackage[authoryear]{natbib}

\pubyear{2019}
\volume{355} 
\jname{The Realm of the Low-Surface-Brightness Universe}
\editors{D. Valls-Gabaud, I. Trujillo \& S. Okamoto, eds.}

\title[The MATLAS project] 
{MATLAS: a deep exploration of the surroundings of massive early-type galaxies}

\author[Pierre-Alain Duc \& MATLAS collaboration]   
{Pierre-Alain Duc$^1$
 \and MATLAS collaboration$^2$}

\affiliation{$^1$Universit\'e de Strasbourg, CNRS, Observatoire astronomique de Strasbourg (ObAS), UMR 7550, 67000 Strasbourg, France \\ email: {\tt pierre-alain.duc@astro.unistra.fr} \\[\affilskip]
$^2$ Project web page: {\tt http://obas-matlas.u-strasbg.fr}}

\begin{document}

\maketitle

\begin{abstract}
The MATLAS project explores the surroundings of a complete sample of nearby massive early-type galaxies with multi-colour deep optical images obtained at the Canada-France Hawaii Telescope. The observing  and data reduction techniques ensured the detection of extended low-surface-brightness (LSB) structures, while the high image quality allowed us to identify associated compact objects such as star clusters. A number of scientific topics are addressed with this data-set that are briefly presented in this review: the study of foreground Galactic cirrus at high spatial resolution, telling us about the ISM structure; the characterisation of collisional debris around the galaxies  (streams, tails, shells, stellar halos), giving hints on their past merger history; the distribution of dwarf galaxy satellites, including the ultra-diffuse ones, together with their globular cluster population, which are additional tracers of the formation and mass assembly of galaxies.    

\end{abstract}

\firstsection 

\section{Introduction}
While the Universe has been scrutinised for decades at all wavelengths, including at exclusive high spatial resolution and depth for some regions, a component of it, ironically located in the vicinity of our Milky-Way, is still largely unexplored. For technical reasons, such as the advent of electronic cameras with initially limited fields of view, luminous but very extended, i.e. diffuse, stellar structures remained for long hidden in the sky background.   
Opening a window on the Low Surface Brightness (LSB) realm, one of the last Terra incognita in our exploration of the Universe (excluding of course its dark components) appears  even more important given that cosmological models predict the presence of a wealth of structures surrounding galaxies, a result of the underlying hierarchical scenario for structure formation. 

As a matter of fact, dramatic advances in LSB studies have recently been made. Extended halos and streams were discovered around the Milky Way  and the Andromeda galaxy using the stellar counts  from ground-based surveys (i.e., SDSS, Pan-STARRs, DES, PandAS/CFHT) or from HST at unprecedented surface brightness limits, confirming the predictions of  numerical simulations.

Beyond the Local Group, identifying individual stars becomes challenging, and the detection of LSB structures mainly relies on the possibility to detect their diffuse light.
The stunning ultra deep images of nearby galaxies obtained by amateur telescopes have largely contributed to convince professional astronomers to use and adapt already-existing facilities (i.e., Burrell telescope, CFHT, VST, Subaru, etc.) or develop their own concepts (Dragonfly, Huntsman) to perform LSB studies.

A number of either targeted observations or larger surveys are currently being carried out by several teams. Their initial results have been presented in a  high number of conferences organised during the last few years that were fully dedicated to the LSB exploration, including this IAU symposium.   Mentioning only the current vitality of this field of research would however be somehow unfair, regarding all the works made during the 20th century in the 50s to 70s with Schmidt photographic plates, which already had the capability to reveal faint extended light, including that being emanated by the so-called Ultra-Diffuse Galaxies.

Among the various studies done with the modern facilities, the originality of the MATLAS survey presented here lies in the high number of galaxies it explores (about 200), the completeness of the observed sample  (using complementary data from the NGVS Survey towards the Virgo cluster), and the high image quality (IQ) of the 
camera (MegaCam on the CFHT) which allowed us to make with the same dataset multiple complementary projects. They are listed in the following sections.

\section{Survey observing strategy and methodology}

The  targets of the MATLAS survey consists of galaxies from the volume limited complete  \atlas\ sample \citep[]{Cappellari11} of nearby (distance below 42 Mpc) massive (absolute K-band magnitude below -21.5) early type galaxies observable from the northern hemisphere ($|\delta-29^\circ|<35^\circ$) and avoiding low Galactic latitudes.
Galaxies too close to bright stars were excluded as well as members of the  Virgo Cluster since they were already observed with MegaCam as part of the ``Next Generation Virgo Cluster Survey (NGVS)'' \citep{Ferrarese12}.

\begin{figure}[h]
\begin{center}
 \includegraphics[width=\textwidth]{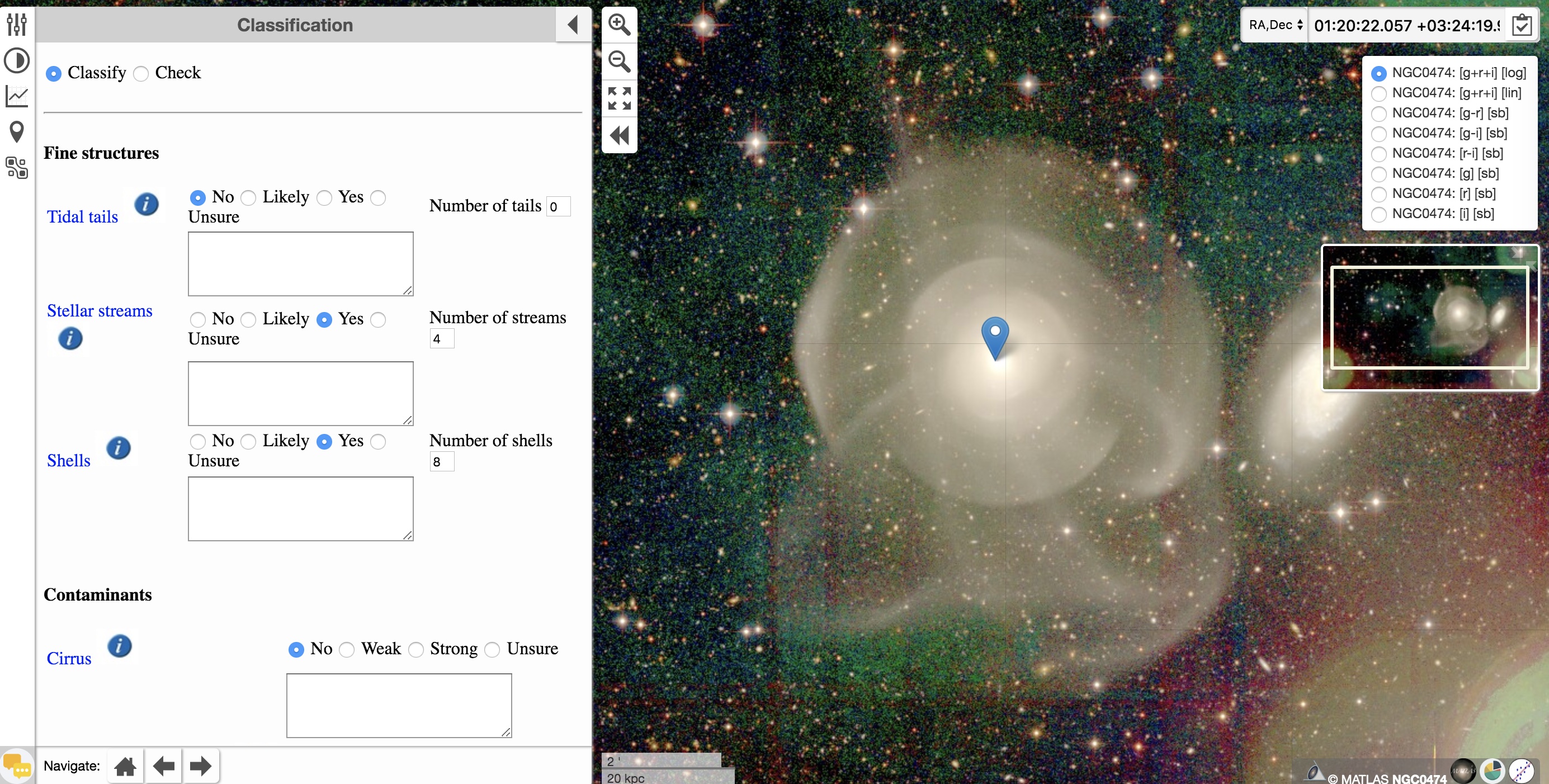} 
 \caption{Web interface, navigation tool and questionnaire used for the exploration  of the MATLAS images and classification of the detected structures.}
  \label{fig:navigator}
\end{center}
\end{figure}

Deep optical images were obtained  for a total 
of 177 galaxies. All but one were observed in both the g' and i' bands; 60\% have also i' band data, and 7 \% have  additional  u* band observations. Total exposure times ranged between 82 minutes (u) 40 minutes (g,r) and 27 minutes (i). The average seeing ranged between 0.65 arcsec (i) and 0.95 arcsec (g). The observing strategy - a sequence consisting of 7 individual exposures with large relative offsets (between 2 and 14 arcmin) and the data reduction technique using Elixir-LSB (sky subtraction, stacking) are fully detailed in \cite{Duc15}. The achieved surface brightness limit is 28.5--29 mag.arcsec$^{-2}$ in the g band.      

Images were then explored and analyzed by the members of the MATLAS team \citep[see details in ][]{Bilek20} using a dedicated web interface, offering  navigation and zoom-in/zoom-out facilities through various types of images for a given target (including composite true color images, color index  and surface brightness maps). The intensity cuts and contrast may be changed online to enhance LSB features (see Fig.\,\ref{fig:navigator}). 
The interface allowed the participant to access a questionnaire  about the presence and properties of stellar structures in and around the target galaxy, about their small and large scale environments and about the presence of contaminants (bright stellar halos, cirrus, artefacts). A visual identification and classification was  used to characterise the fine structures around the ETGs. 

Light and colour profiles of the galaxies up to large effective radii (typically 10 Re) were obtained with ellipse fitting techniques, in images that had previously been corrected from PSF effects.  A deconvolution technique presented in \cite{Karabal17} was used for this correction.   Applying it turned out to be fundamental  as the outer halos of the bright galaxies on MegaCam images  are contaminated by light coming from the central parts of the galaxies because of internal reflections in the camera. The PSF effect which varies from one band to the other, and is most pronounced in the r band for MegaCam, generates an artificial reddening in the g-r colour profiles of some galaxies.  

The census of dwarf galaxies in the field-of-view of each target (about 1 square degree) was made with a set of semi-automatic algorithms applied on cleaned images, followed by a visual confirmation and classification by the team members. The technique is described in details in \cite{Habas20}.   

Finally, globular cluster (GC) candidates in all fields were identified using the available colour information  and a proxy of the size obtained comparing magnitudes in two apertures. Further details on this photometric method are given in \cite{Lim17}. The GC spectroscopic confirmation  could be achieved in a few specific fields for which observations with the MUSE integral field spectrograph  on the VLT have been obtained, e.g. towards the shell galaxy NGC~474 \citep{Fensch20} and a few UDG candidates \citep{Mueller20}. The catalog of GC candidates in the MATLAS fields contains about 30 000 objects (about 200 per field).

\section{Getting through the foreground emission: cirrus}

\begin{figure}[h]
\begin{center}
 \includegraphics[width=\textwidth]{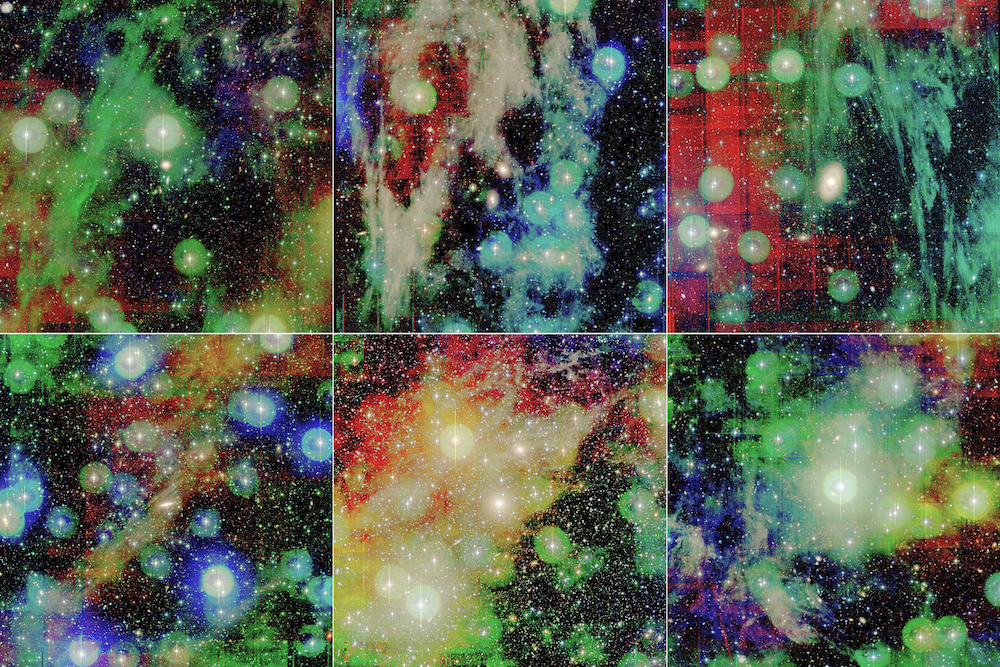} 
 \caption{Examples of fields strongly polluted by Galactic cirrus. The displayed composite (g+r+i) images have a size of 1 square degree. Also note the 7' arcmin diameter reflection halos around the bright stars.}
  \label{fig:cirrus}
\end{center}
\end{figure}

As illustrated in Fig.\,\ref{fig:cirrus}, at the surface brightness limit of MATLAS, some fields are particularly crowded with the extended reflection halos of bright stars  and emission from Galactic cirrus, leaving only small unpolluted regions  to study the LSB structures around the target galaxies. 
The light that scatters in  dust clouds shows up as either very uniform smudges located around bright stars or  stripped structures, sharing a common orientation. They do not seem to have specific colors at least with the set of filters used for MATLAS. Their filamentary structures  actually mimic those of stellar streams and may be confused with them (see an example in \citealp{Duc18}). However a comparison to  data at other wavelengths, especially in the mid--infrared (for instance with the WISE all-sky survey) or far infrared regime (for instance with Planck), sensitive to the dust thermal emission, leaves little  doubts on the nature of the Galactic nature of the emission.  

Cirrus contamination becomes more important towards  the Galactic plane, but it may be present even at high Galactic latitude. 
Indeed, as discussed  in \cite{Miville16} the culprits may be dust clouds located in the vicinity of the Sun, just a few hundred  parsecs away. 

Cirrus emission might be the main limiting issue for the exploration of the LSB Universe, especially if one wishes to reach very low surface brightness limits, for instance from Space observatories. Looking at the maps generated  by the Planck mission, one may wonder whether the whole sky has real clean extragalactic windows for LSB studies. This calls for attempting to subtract this foreground component as done for the CMB analysis.
Because cirrus does not seem to have a specific color -- the g-r color index varies along the structures -- , this challenging task will likely require a compilation of the complete spectral energy distribution of the data, from the far UV to the far IR, and advanced techniques of data processing.   Meanwhile, the current deep surveys tend to avoid regions of heavy pollution by  cirrus.
 
 Instead, one  may fully exploit their presence and nearby location, to carry out studies of the interstellar medium at unprecedented spatial resolution.  \cite{Miville16} illustrate some of the scientific questions that may be  addressed determining the power spectrum of the cirrus emission from the largest scales given by Planck to the smallest ones probed by the optical images. They involve testing  the predictions on the turbulence cascade in the ISM.

\section{Exploring the surrounding of galaxies: fine structures and halos}

Determining the fraction of galaxies exhibiting in their surroundings tidal features (also referred as fine structures, FS) was one of the original motivations of the project. FS  include  narrow {\em streams} -- tidal features made from stars pulled out from a disrupted low-mass companion, diffuse {\em tails} -- antennae or plume-like structures from the primary galaxy tracing an on-going or past major merger,  or {\em shells} -- concentric sharp-edge stellar structures, due to specific types of collisions. Such tracers of the past mass assembly of galaxies may be destroyed by successive collisions or gradually dilute with time scales depending on the FS type, as shown by \cite{Mancillas19} using numerical simulations.

\begin{figure}
\begin{center}
 \includegraphics[width=\textwidth]{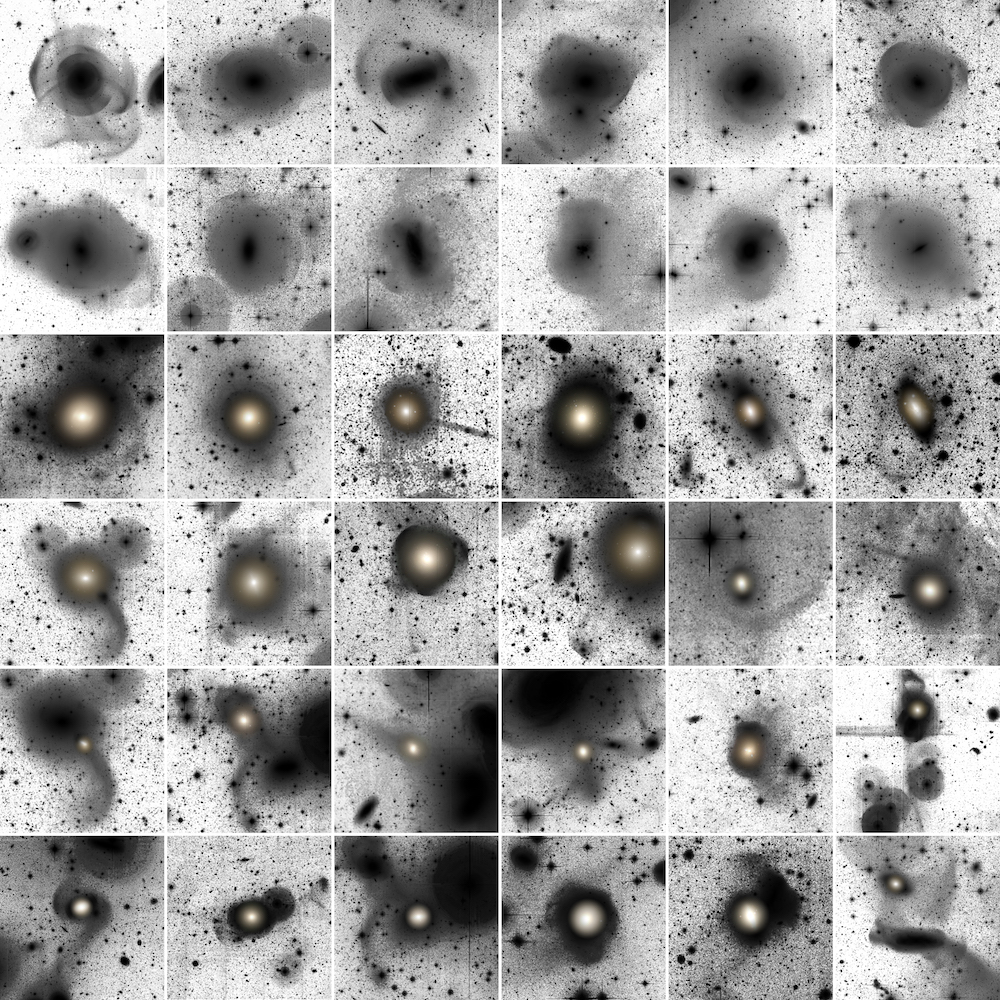} 
 \caption{Various types of identified and classified fine structures in the MATLAS survey. Top first rows: shells, middle rows: streams, bottom rows: tails. The images consist of co-added g+r frames, with, for the stream and tail categories,  the true color images superimposed towards the central region. This helps to identify the primary target galaxy. See more images in \cite{Bilek20}.}
  \label{fig:finestructures}
\end{center}
\end{figure}

Therefore, as each type of  features may tell a different story about the nature of the past merging history  of a galaxy  (minor vs major merger) and of its progenitors  (late or early type galaxies), it was important to not only count the FS, but also to determine their morphology and distribute them in different classes. Currently, this is most easily  done  with a visual inspection of the images, preferentially by expert eyes -- a conclusion from the various tests made by \cite{Bilek20} in order to estimate the reliability of the manual classification. 
Once the debris have dynamically relaxed  or evaporated, extended stellar halo can still bring information about the old merger events. 
Therefore members of the MATLAS team were invited to identify, classify and count the  fine structures, but also to characterise the shape of the extended stellar halos. Examples of ETGs classified as a function of their main types of collisional debris are presented in Fig.\,\ref{fig:finestructures}.

With such a statistical database, we could compute the fractions of galaxies having a specific type of features and correlate them with the internal properties of the target galaxies (mass, internal dynamics, presence or kinematical substructures such as KDCs) and with their environment, probed by a tracer of the local density of galaxies. This was possible thanks to the wealth of complementary multi-wavelength data provied by the \atlas\ project. The main results, presented in details in \cite{Bilek20} and  B\'ilek et al. (in prep.), are summarised here. They refer to only ETGs located in the MATLAS fields, thus ignoring the galaxies from the \atlas\ sample located in the densest environments, i.e. in the Virgo cluster.
\begin{itemize}
    \item Globally about 40\% of all ETGs in the sample show evidence of tidal features or asymmetric stellar halos.
    \item The fraction of galaxies showing either likely or secure streams, tails and shells is about the same, around 15\%.
    \item The fraction is mass dependent. For instance, above the mass of $10^{11}$\,M$_\odot$, the fraction of fully relaxed objects drops to around 40\%. Streams and shells are twice more frequent above that mass.
    \item Slow rotators have on average 1.3 more tidal features than fast rotators.
    \item  All these statistics seem to only weakly vary with the environment as traced by the galaxy density, but a more detailed analysis is needed to ensure this.
\end{itemize}

\section{Exploring the surroundings of galaxies: dwarf satellites and globular clusters}

\begin{figure}[h]
\begin{center}
 \includegraphics[width=\textwidth]{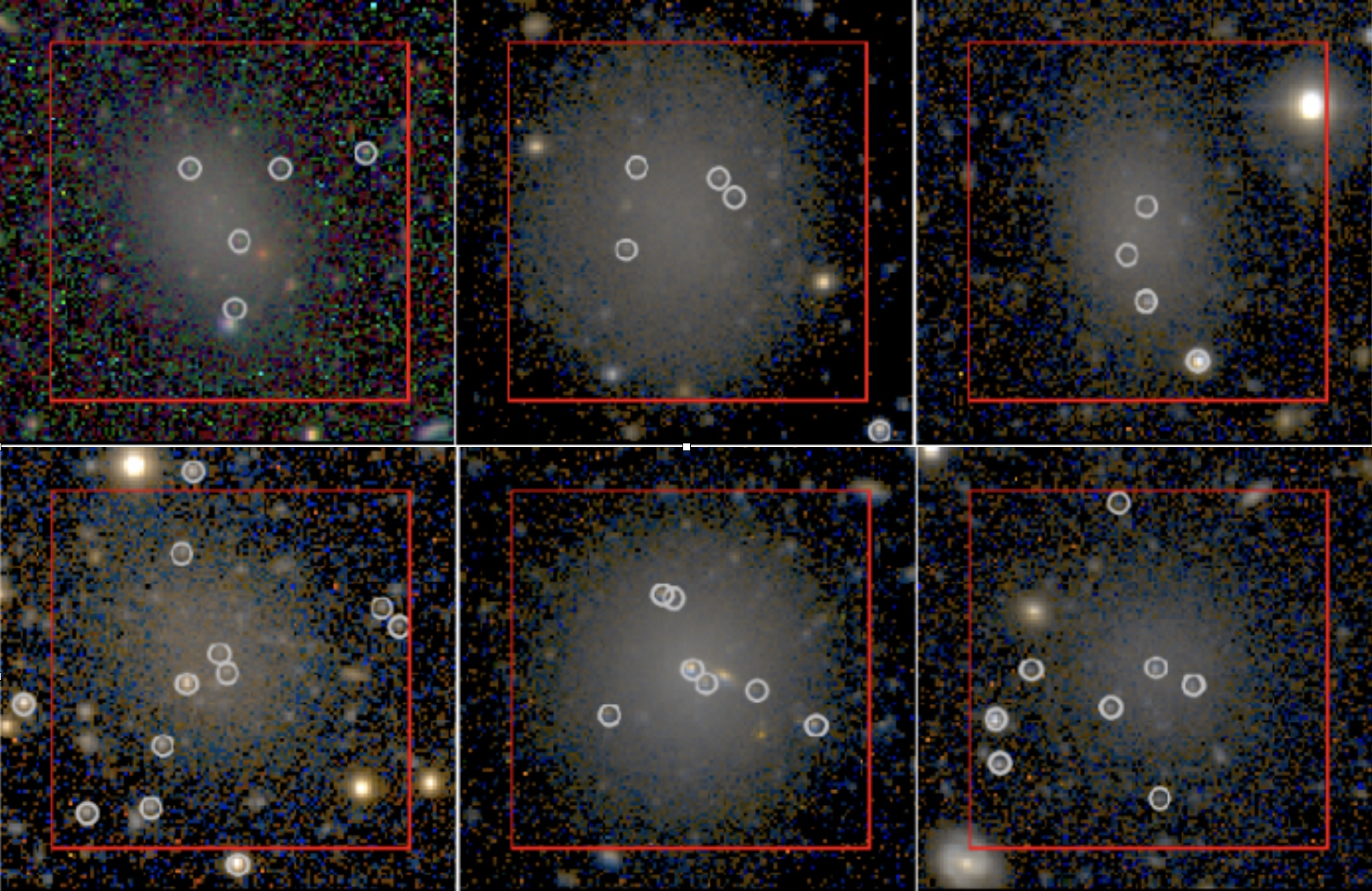} 
 \caption{Dwarf satellite candidates in the field of MATLAS galaxies, including UDGs. They show an excess of globular cluster candidates  identified by the circles.}
  \label{fig:dwarfs}
\end{center}
\end{figure}

Besides, beyond the collisional debris, stellar objects such as dwarf galaxies and globular clusters may be used to probe the large-scale environment of galaxies, and get further clues on their structural properties and formation mechanism.

A total of 2210 dwarf galaxy candidates have been identified in the MATLAS fields,  covering together a total area of 142 deg$^2$ \citep[]{Habas20}. Obviously the absence of a distance indicator for most of them prevents us from ensuring their nature as dwarf objects and physical association with the massive galaxies in the field. This could however successfully be tested for about 10\% of them with available velocity measurements from archival data (SDSS or WSRT HI data from the \atlas\ survey). 
About 75\% of the dwarf candidates are early-type and may be classified as dEs. They are thus abundant
even outside  galaxy clusters  where they are usually detected. One fourth of them are nucleated with a fraction which increases in  denser environments, a trend also found in previous surveys. 
Assuming  that their distances correspond to that of the massive host in the field, the dwarf candidates follow the scaling relations determined previously (effective radius vs absolute magnitude vs surface brightness).

This is a further indication about  the reliability of the selection process of the dwarf candidates.  Among  them, about 90 (4\% in our sample, or 0.6  per square degree) have an effective radius above 1.5 kpc and central surface brightness below 24 mag/arcsec$^2$ and would qualify as Ultra Diffuse Galaxies (UDGs). Whether UDGs make or not a genuine class of objects remains highly controversial. Nonetheless it is interesting to note that while the vast majority of UDGs have yet been detected in clusters, we do find some of them in the environments probed by MATLAS, i.e. in the vicinity of ETGs in isolation or groups. 

A tiny fraction (1\%) of the dwarf candidates -- 4\% considering only those qualifying as UDGs -  exhibit an excess of globular cluster candidates (more than 4 GCs, when 1-2 are expected given the mass of the host). Examples are shown in  Fig.\,\ref{fig:dwarfs}.   Understanding how and when these low mass objects managed to form or collect this population of GCs is the focus of on-going spectroscopic follow-up using integral field spectrometers and HST observations. The GC  excess was spectroscopically  confirmed for MATLAS-2019 \citep{Mueller20}.

\section{Getting further: challenges and prospects}

\begin{figure}[h]
\begin{center}
 \includegraphics[width=\textwidth]{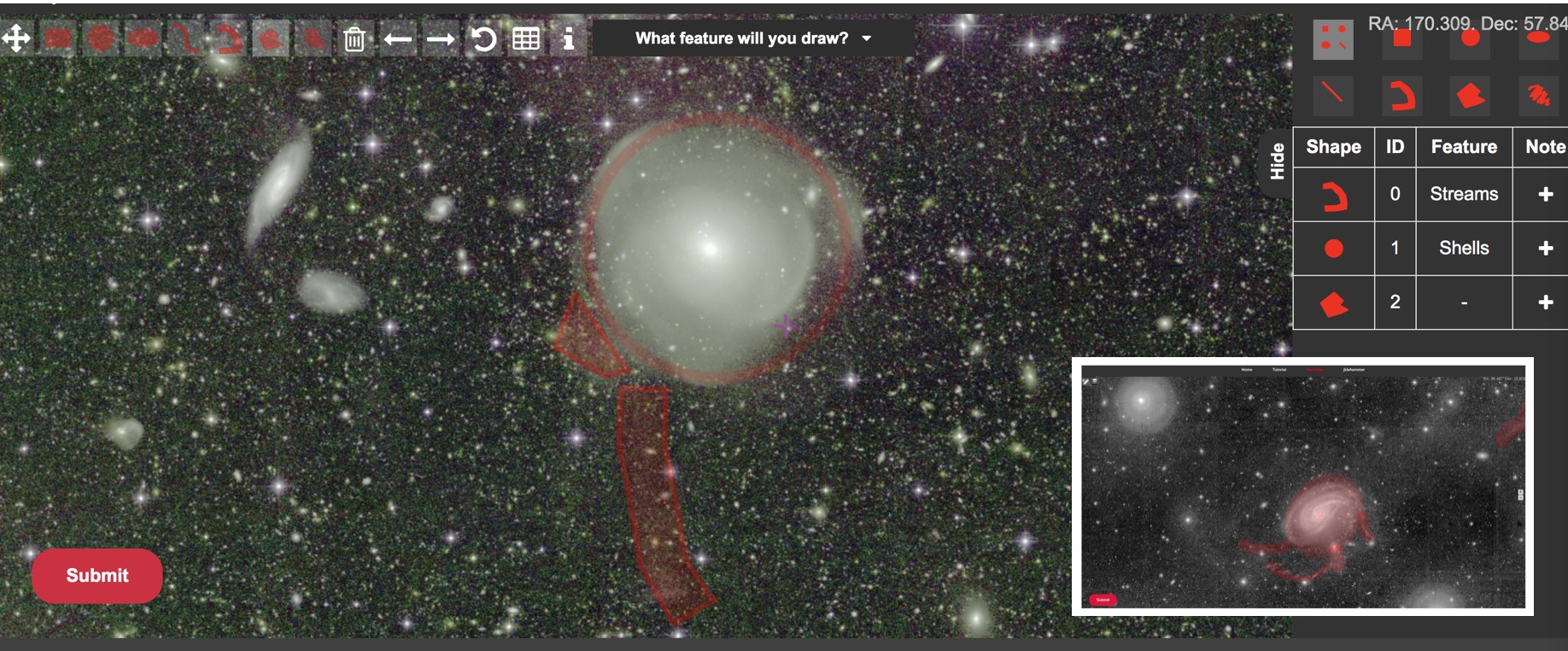} 
 \caption{Annotation interface based on the CDS Aladin lite tool}
  \label{fig:annotate}
\end{center}
\end{figure}

Many of the results presented above rely on visual identifications  and classifications. 
This could be achieved by a small group of people because of the limited number of fields  but will be unpractical for the on-going (e.g. CFIS/Unions on the CFHT) or future (LSST, Euclid)  surveys that cover  much larger areas (several thousands of square degrees). The use AI  techniques, such as deep learning with a training based on visual classifications by large groups of people or data generated by numerical simulations, will soon be unavoidable.

Getting more quantitative results would also be an asset. This implies obtaining a reliable photometry of the faint collisional debris and, before that, delineating them precisely. For this purpose, we have developed a web interface  to manually and quickly mark the fine structures (Richards et al., in prep.  and Fig.\,\ref{fig:annotate}). Alternatively one may consider using  automatic detection of tidal features with algorithms such as {\em NoiseChisel} \citep[]{Akhlaghi15}. Further developments are however needed to improve the segmentation process necessary to separate the collisional debris from the host galaxy. 

Given the tremendous  progresses made in AI and advanced data processing techniques, there is little doubt that the next IAU symposium dedicated to the exploration of the LSB Universe will present works fully relying on  automatic  techniques for the  identification and classification of structures.

\acknowledgements 
This contribution is given on behalf of the full MATLAS team. Special thanks to Michal B\'ilek, Jean-Charles Cuillandre, Stephen Gwyn, Rebecca Habas, Sungsoon Lim, Francine Marleau, Marc-Antoine Miville-Desch\^enes, Oliver M{\"u}ller and  Mustafa Yildiz who have directly contributed to the results highlighted in this paper. The full list of team members, and image navigation tool,  are available at the MATLAS project web site: http://obas-matlas.u-strasbg.fr

\begin{discussion}

\discuss{C. Mihos}{What is the surface brightness depth reached by MUSE on the shells of NGC 474\,?}
\discuss{P.A. Duc}{It is about 25-26 mag/arcsec${^2}$ in the g-band}

\discuss{B. Koribalski}{Can you tell us more about features detected by amateurs, but not seen at CFHT\,?}
\discuss{P.A. Duc}{These are the features that are lost behind the prominent reflections halos of the MegaCam camera. Note however that in general the CFHT MATLAS images are deeper that those obtained by amateurs, and exhibit features not seen by them. Thus both approaches are complementary.  }

\end{discussion}

\end{document}